\documentclass[useAMS,usenatbib]{mn2e}
\usepackage{graphicx,fleqn,fix2col,rotating}
\DeclareGraphicsExtensions{.eps,.ps,.eps.gz,.ps.gz,.eps.Z}
\title[Transmission photometry of WASP-12b]{Transmission photometry of WASP-12b: simultaneous measurement of the planetary radius in three bands}

\author[C.M.~Copperwheat et al.]{C.M.~Copperwheat$^{1,2}$\thanks{c.m.copperwheat@ljmu.ac.uk}, P.J.~Wheatley$^{2}$\thanks{p.j.wheatley@warwick.ac.uk}, J.~Southworth$^{3}$, J.~Bento$^{4}$, \newauthor T.R.~Marsh$^{2}$, V.S.~Dhillon$^{5}$, J.J.~Fortney$^{6}$, S.P.~Littlefair$^{5}$ and R.~Hickman$^{2}$\\\\
$^{1}$ Astrophysics Research Institute, Liverpool John Moores University, IC2, Liverpool Science Park, 146 Brownlow Hill,
Liverpool, L3 5RF, UK\\
$^{2}$ Department of Physics, University of Warwick, Coventry, CV4 7AL, UK\\
$^{3}$ Astrophysics Group, Keele University, Keele, Staffordshire, ST5 5BG, UK\\
$^{4}$ Department of Physics and Astronomy, Macquarie University, NSW 2109, Australia\\
$^{5}$ Department of Physics and Astronomy, University of Sheffield, Sheffield, S3 7RH, UK\\
$^{6}$ Department of Astronomy and Astrophysics, University of California, Santa Cruz, CA 95064, USA\\
}

\date{Received: }

\begin{document}

\newcommand{\dg} {^{\circ}}
\outer\def\gtae {$\buildrel {\lower3pt\hbox{$>$}} \over
{\lower2pt\hbox{$\sim$}} $}
\outer\def\ltae {$\buildrel {\lower3pt\hbox{$<$}} \over
{\lower2pt\hbox{$\sim$}} $}
\newcommand{\ergscm} {erg s$^{-1}$ cm$^{-2}$}
\newcommand{\ergss} {erg s$^{-1}$}
\newcommand{\ergsd} {erg s$^{-1}$ $d^{2}_{100}$}
\newcommand{\pcmsq} {cm$^{-2}$}
\newcommand{\ros} {{\it ROSAT}}
\newcommand{\xmm} {\mbox{{\it XMM-Newton}}}
\newcommand{\exo} {{\it EXOSAT}}
\newcommand{\sax} {{\it BeppoSAX}}
\newcommand{\chandra} {{\it Chandra}}
\newcommand{\hst} {{\it HST}}
\newcommand{\subaru} {{\it Subaru}}
\def\rchi{{${\chi}_{\nu}^{2}$}}
\newcommand{\Msun} {$M_{\odot}$}
\newcommand{\Mwd} {$M_{wd}$}
\newcommand{\Mbh} {$M_{\bullet}$}
\newcommand{\Lsun} {$L_{\odot}$}
\newcommand{\Rsun} {$R_{\odot}$}
\newcommand{\Zsun} {$Z_{\odot}$}
\newcommand{\Mjup} {$M_{J}$}
\newcommand{\Rjup} {$R_{J}$}

\def\Mdot{\hbox{$\dot M$}}
\def\mdot{\hbox{$\dot m$}}
\def\mincir{\raise -2.truept\hbox{\rlap{\hbox{$\sim$}}\raise5.truept
\hbox{$<$}\ }}
\def\magcir{\raise -4.truept\hbox{\rlap{\hbox{$\sim$}}\raise5.truept
\hbox{$>$}\ }}
\newcommand{\mnras} {MNRAS}
\newcommand{\aap} {A\&A}
\newcommand{\apj} {ApJ}
\newcommand{\apjl} {ApJL}
\newcommand{\apjs} {ApJS}
\newcommand{\aj} {AJ}
\newcommand{\pasp} {PASP}
\newcommand{\aaps} {AAPS}
\newcommand{\apss} {Ap\&SS}
\newcommand{\araa} {ARAA}
\newcommand{\nat} {Nature}
\newcommand{\pasj} {PASJ}
\newcommand{\ha}{\hbox{$\hbox{H}\alpha$}}
\newcommand{\hb}{\hbox{$\hbox{H}\beta$}}
\newcommand{\hg}{\hbox{$\hbox{H}\gamma$}}
\newcommand{\heii}{\hbox{$\hbox{He\,{\sc ii}\,$\lambda$4686\,\AA}$}}
\newcommand{\hei}{\hbox{$\hbox{He\,{\sc i}\,$\lambda$4472\,\AA}$}}

\maketitle

\begin{abstract} 
Transmission spectroscopy  has been successfully used from both the ground and in space to characterise the atmospheres of transiting exoplanets. This technique is challenging from the ground because ground-based spectrographs tend not to be designed to be photometrically stable, and effects such as variable slit losses cause significant systematic uncertainties. An alternative approach is to use simultaneous photometric observations in multiple wavebands to determine wavelength dependent transit depth differences. We report an application of this technique to one of the hottest known exoplanets, WASP-12b, using the triple-beam camera ULTRACAM. We obtained simultaneous light curves in Sloan $u$', and two narrow band filters centered on $4169$\AA \ and $6010$\AA, with FWHMs $52$\AA \ and $118$\AA \ respectively. We fit these light curves with a photometric model and determine the planetary radius in the three different bands. Our data show no evidence for a difference in planetary radius over the wavelength range we study, and are consistent with an atmosphere that is dominated by Rayleigh scattering from a high altitude haze, as well as more complicated atmosphere models which include the effects of molecules such as TiO. Our planetary radius measurements have an average precision of $2.6$ per cent, compared to the $\sim$$1.4$ -- $2.4$ per cent radius differences predicted by the models over this wavelength range. We also find a consistent time of ingress and egress across our three wavebands, in contrast to the early ingress which has been reported for this system at shorter wavelengths.
\end{abstract}

\begin{keywords}
stars: planetary systems -- stars: fundamental parameters -- stars: individual:WASP-12
\end{keywords}
%%%%%%%%%%%%%%%%%%%%%%%%%% Begin Section 1 %%%%%%%%%%%%%%%%%%%%%%%%%%%%%
\section{INTRODUCTION}  
\label{sec:intro}

The study of extrasolar planets (exoplanets) is one of the most rapidly advancing areas of modern astronomy. Around a third of known exoplanets have been observed to transit their host star, which provides a significant advantage for the elucidation of their nature, since not only does the edge-on viewing geometry remove the uncertainty in the inclination, but the light curve shape itself is dependent on the sizes of the star and planetary components. The resulting mass-radius relation (e.g., \citealt{Pollacco08}) provides the basis for understanding the structure, composition and evolution of the exoplanet population.

In addition to the mass-radius relation, transit observations provide the means to probe a variety of other physical properties of exoplanets. Observations over a range of wavelengths allow us to investigate the composition of the planetary atmospheres, since opacity sources raise the altitude of the photosphere at certain wavelengths and thus have a measurable effect on the planetary radius at those wavelengths. This {\it transmission spectroscopy} technique was applied to HD209458b using the Space Telescope Imaging Spectrograph (STIS) on the Hubble Space Telescope ({\it HST}), resulting in the detection of atomic sodium in the planetary atmosphere \citep{Charbonneau02}. Further STIS observations were obtained by \citet{Knutson07}, and a simultaneous fit of ten photometric bandpasses yielded even more precise determinations of the planetary radius and orbital inclination. The discovery of atomic sodium in HD209458b was followed by a number of detections of other species in the upper atmosphere of exoplanets, using {\it HST} and the transmission spectroscopy technique (e.g., \citealt{Vidal03,Vidal04,Lecavelier10}). The atmosphere of HD209458b has since been detected at low spectral resolution across the entire optical waveband, also using {\it HST} \citep{Sing08,Lecavelier08}. The resulting spectrum is dominated by a blue broadband opacity source, interpreted as Rayleigh scattering by H$_2$, as well as a Stark-broadened component of the Na {\sevensize I} line. A low-resolution {\it HST} spectrum of HD189733b also showed an atmospheric haze, in this case a scattering continuum from a haze of submicrometre particles \citep{Pont08,Sing11}.

There have been a number of successful applications of the technique of transmission spectroscopy from the ground. This has resulted in the confirmation of sodium absorption \citep{Redfield08, Snellen08}, and the detection of H$\alpha$ \citep{Jensen12} in the atmospheres of both HD~189733b and HD~209458b. Other recent examples are the detection of sodium in WASP-17b \citep{Zhou12} and sodium and potassium in XO-2b \citep{Sing12}. However, transmission spectroscopy has been more challenging from the ground because ground-based spectrographs tend not to be designed to be photometrically stable, and effects such as variable slit losses cause significant systematic uncertainties. The best precision is currently achieved using multi-object spectrographs (e.g. \citealt{Bean10}). Simultaneous observations of the target and multiple comparison stars allows atmospheric variations to be accounted for, and wide entrance apertures minimise slit losses.

An alternative approach from the ground is to take simultaneous photometric observations in multiple bands, and then fit the transit light curve in each band in order to detect differences in the planetary radius with wavelength. This avoids some of the complications asociated with spectroscopy, at the price of spectral resolution. However, since observations have shown many planetary atmospheres are dominated by broadband features, photometric observations can still be an effective means of distinguishing between models. A recent example of this technique was presented in \citet{Southworth12a}, with simultaneous observations in Str\"omgren $u$, Gunn $g$ and $r$, and Johnson $I$ filters of a transit in the HAT-P-5 system. They found the planetary radius to be larger in $u$, but could not rule out systematic errors as the cause of this difference.

In 2009  we obtained a transit of WASP-12b with the triple-beam high speed CCD camera ULTRACAM \citep{Dhillon07} mounted on the $4.2$m William Herschel Telescope (WHT). WASP-12b was discovered by the SuperWASP collaboration \citep{Hebb09}, and at the time of its discovery (and our follow-up observations) it was the hottest transiting exoplanet known, making it an excellent target with which to test atmosphere models. In addition, recent {\it HST} Cosmic Origins Spectrograph (COS) observations of this star were interpreted by \citet{Fossati10} to show an early ingress of the exoplanet at near-ultraviolet wavelengths. This has been modelled as due to the Roche lobe overflow of material from the star \citep{Lai10} or alternatively due to a bow shock of coronal material around the magnetosphere of the planet \citep{Llama11}. Simultaneous observations in multiple optical wavebands provides a test of these models.

\section{OBSERVATIONS AND REDUCTION}
\label{sec:obs}

On the night of 04 January 2009 we observed WASP-12b for $10$ hours, beginning at $19$:$34$ UT and covering one transit, using the fast photometer ULTRACAM mounted on the WHT. ULTRACAM is a triple beam camera, taking observations simultaneously in three bands. Typically it is used with Sloan broadband filters, but we chose to use narrow-band filters in the green and red arms in order to minimise atmospheric trends, as we will discuss in Section \ref{sec:atmos}. In the green arm we used a filter with a central wavelength of $4169$\AA \ and a FWHM of $52$\AA. In the red arm we used a filter with a central wavelength of $6010$\AA \ and a FWHM of $118$\AA. We retained the Sloan $u$'-band filter in the blue arm (central wavelength $3557$\AA, FWHM $599$\AA). An additional motivation for this choice of filters was the theoretical work of \citet{Fortney08}, which presented two classes of model atmospheres, one of which features appreciable opacity at certain wavelengths due to TiO and VO. In order to try and detect this effect we chose narrow band filters in which the disparity between the observed planetary radii is predicted to be at its greatest (as shown in figure 11 of \citealt{Fortney08}). 

For the majority of the observation we used a $5.8$s exposure time. The images were unbinned, and the full frame of the CCD was used with the fast readout mode. The dead time between exposures for ULTRACAM is $\sim$$25$ms. The night was photometric with acceptable ($\sim$$1$'' on average) seeing, although the seeing was quite variable on short timescales, particularly in the early part of the night. As has become fairly standard in observations of planetary transits we applied a modest defocus of the telescope such that the FWHM was $\sim$$2$''. Aside from the possibility of saturation when observing a bright star like WASP-12 ($B = 12.1$), the purpose of the defocus is to reduce the systematic uncertainty introduced by the flatfielding, since by increasing the FWHM, the contribution of any one individual pixel to the overall measured flux is reduced. This uncertainty is potentially quite important, and for the same reason the star should ideally be held in the same position on the chip over the course of the observation. We relied on the autoguider and did see a drift in pixel position over the course of our observation of $10$ -- $15$ pixels ($3$ -- $4.5$''), but we find no evidence that this drift introduces any systematic trend into our data.

These data were reduced with aperture photometry using the ULTRACAM pipeline software\footnote{http://deneb.astro.warwick.ac.uk/phsaap/\\software/ultracam/html/index.html}, with debiassing, flatfielding and sky background subtraction performed in the standard way. Flatfielding was performed using $\sim$$500$ sky flats which were obtained on the same night as the transit observation. The source flux was determined using a variable aperture, whereby the radius of the aperture is scaled according to the FWHM. The scaling factor for each band was determined empirically by reducing each dataset with a range of possible factors, and choosing the ones which produced the light curves with the lowest scatter in a small section of the out-of-transit data. The use of a variable aperture can potentially cause problems when applied to defocused observations, due to the reduction routine failing to determine correct centroid and FWHM values. However, in a case such as this where we have only applied a small defocus we found the variable aperture method to be somewhat superior to using a fixed aperture by comparing the S/N obtained in each case. The centroiding was performed using a Moffat profile fit. As well as the target we also selected and reduced seven comparison stars (listed in Table \ref{tab:comps}) which we used to correct short-timescale transparency variations in our target light curves, as well as the longer timescale atmospheric trends. This process is described in detail in Section \ref{sec:atmos}. The positions of these comparison apertures were fixed relative to the position of the primary aperture, which helps the centroiding for these fainter stars. Where appropriate we used the ULTRACAM software to mask fainter stars which overlap with the sky apertures. `Lucky Imaging' observations have detected a faint M-dwarf star approximately $1$'' from Wasp-12b \citep{Bergfors13}. This is too close to the target to mask, but likely contributed a negligible amount of flux at the wavelength range we study. For all data we convert the MJD times to the barycentric dynamical timescale (TDB), correcting for light travel times. We did not use the section of data obtained at the very end of the night at high ($>2$) airmasses, as we found no benefit to the parameter determinations when we included these data.

\begin{table}
\caption{Coordinates of the comparison stars we used for the differential photometry.}
\label{tab:comps}
\begin{center}
\begin{tabular}{lllll}
\# & Name & RA (hms) & Dec (dms)\\
\hline
1 &TYC 1891-326-1&$06:30:39.8$ &$+29:37:40.4$\\
2 &TYC 1891-324-1&$06:30:48.3$ &$+29:39:35.9$\\
3 &TYC 1891-38-1 &$06:30:49.9$ &$+29:38:54.7$\\
4 &&$06:30:46.40$ &$+29:41:12.8$\\
5 &&$06:30:45.81$ &$+29:41:28.0$\\
6 &&$06:30:32.23$ &$+29:37:34.7$\\
7 &&$06:30:43.23$ &$+29:37:53.7$\\
\hline
\end{tabular}
\end{center}
\end{table}

\section{LIGHT CURVES AND MODEL FITTING}
\label{sec:fitting}

\subsection{Atmospheric effects}
\label{sec:atmos}

In differential photometry both the target star and a nearby, photometrically stable comparison star are reduced simultaneously using the same sized apertures. The light curve of the target is then divided by the light curve of the comparison. If the comparison is close enough to the target this should remove the short ($\sim$exposure time) timescale variation due to changes in transparency. It should also remove any long timescale trends in the data. The most significant of these is the variation in the target and comparison count rate due to the changing airmass. The response of the star to changing airmass varies with colour, and so to optimise this correction the comparison star is required to be very close in colour to the target. In the WASP-12b field, we found the brighter comparison stars are not a sufficiently good colour match to remove the airmass trend at the level of photometric precision required ($\sim$$1$ per cent), and so when we divided through by the comparison star there were residual trends. We attempted to minimise this effect by using narrow band filters in the red and green arms. If the filter is narrow enough such that the stellar spectrum is essentially flat over the wavelength range sampled, then the colour difference between the target and comparison will not matter and the light curve should be flattened via the differential photometry without resorting to any further manipulation. This choice came at the price of a reduced count rate in these arms, and in practice we found there was still a residual slope in the differential light curve, albeit much smaller than we find in the blue arm (in which we retained the broad band $u$' filter). There may be other contributing factors to the long-term trends, for example we considered the effect of telescope vignetting: if the two stars are at different positions on the CCD from the rotation axis of the telescope, then as the observation proceeds the rotation of the field may result in an appreciable trend. This is distinct from the instrumental vignetting, which should not change over the course of the observation as long as the star is maintained in the same position on the CCD. However, the fact that the use of narrow band filters offers a significant improvement implies that the colour difference between the target and comparison star is the dominant factor.

We found that the combination of multiple companion stars provided a superior correction of the atmospheric trends compared to using a single star. Accordingly, we selected the seven stars listed
in Table \ref{tab:comps}. These are stars which we found to be photometrically stable, which were bright in all three arms and did not coincide with hot pixels or bad columns on the ULTRACAM CCDs.
Ideally we would use a comparison star which is brighter than the target, but unfortunately WASP-12 is the brightest star in the field of view, although comparisons $1$ and $2$ are close in brightness in the green and red arms. We averaged these seven stars together and divided the target light curves by this average. The price of this superior correction for the atmospheric trends was a reduced signal-to-noise. The trend was not removed completely -- potentially further improvement could be found by including even more comparison stars, but the remaining stars in the field of view are significantly fainter, and so the subsequent reduction in signal-to-noise is too great. We preferred to fit this residual trend as a component in our model, and discuss this in more detail in Section \ref{sec:model} along with the other model components.

In many analyses of exoplanet light curves obtained from the ground, the light curves are flattened with a polynomial which has been fitted to the out-of-transit data, before performing the model fit through which the physical parameters are determined (e.g. \citealt{Southworth09}). The problem with this is that it is often unclear if uncertainty introduced by the initial fit is reflected in the final parameter determinations. Choice of polynomial can also dramatically affect the transit depth, and the sections of out-of-transit data used for the polynomial are often quite small, even if a complete night is spent on target. A polynomial correction outside of the main model fit is particularly unwise in a case such as this in which we are comparing transit depths in different light curves. Potentially some of the inconsistencies in transit depth determinations which can be found in the literature are due to this practice. 

We also considered more complicated approaches to the problem of atmospheric correction. Instead of making an unweighted average of comparison stars, or an average weighted by luminosity, one approach is to have the weights as free parameters and fit them so as to minimise the out-of-transit trends. Another approach is, rather than dividing by the average of the comparisons stars, to instead fit a high order ($10+$) polynomial to each comparison light curve. The average of these polynomials is then used to remove the long term variation from the brightest comparison, and the target light curve is then divided by the corrected light curve of this single comparison to remove the short-term variation. This method should correct the trends in the same way but without any reduction to the signal-to-noise. We tried this approach, but found that when we fitted these light curves with our model the uncertainties on the resultant parameter determinations showed no significant improvement, implying it is systematic factors rather than the photon noise which dominates. We chose therefore for clarity to use the simpler method of averaging the comparison stars, as described above.

\subsection{The light curve model}
\label{sec:model}

We modelled the planetary transit using the {\sevensize LCURVE} light curve fitting code written by T. Marsh, which is described in detail in the appendix of \citet{Copperwheat10}. This code was originally designed to model the light curves of eclipsing cataclysmic variables. We used a modified version of this code in which the accretion disc and bright spot components are removed, and the secondary component has a spherical rather than Roche-lobe filling geometry. Modified in this way the {\sevensize LCURVE} code can be used to describe a detached eclipsing binary or, by setting the flux contribution from the secondary to zero, a transiting exoplanet. The parameters of the model in this case are the mass ratio $q$ and the inclination $i$, the radii of the star and the planet $R_\ast$ and $R_{pl}$, which we express in terms of the orbital separation $a$, the time of mid-transit $t_0$, the orbital period $P$ and the stellar limb darkening coefficients. We also accounted for the residual airmass trend in our data by multiplying the model by 
\begin{eqnarray}
1+t(A+Bt)
\end{eqnarray}
where $t$ is the time scaled such that it varies from $-1$ to $1$ from the beginning to the end of the data, and $A$ and $B$ are additional parameters in the fit.

\subsection{The limb darkening coefficients}
\label{sec:limb}

Our parameter of interest is the planetary radius, $R_{pl}$, which is determined by the depth of the transit. The stellar limb darkening is correlated with this parameter, and so both our choice of limb darkening law and our choice of starting values are important. \citet{Southworth08} discussed the various limb darkening laws as applied to exoplanet transits in detail. He emphasised the importance of using a nonlinear law but noted that using a law with more than two coefficients offered no significant improvement using ground-based data of contemporary quality. Given the importance of this model component, we chose to fit the data twice using two different laws: the polynomial (quadratic) law
\begin{equation} 
{{I(\mu)}\over{I(1)}} = 1 - u_{1 \ast}(1-\mu) - u_{2 \ast}(1-\mu)^2
\end{equation}
and the two-coefficient form of the non-linear law first presented in \citet{Claret00}
\begin{equation}
{{I(\mu)}\over{I(1)}} = 1 - u_{1 \ast}(1-\mu^{1/2}) - u_{2 \ast}(1-\mu).
\end{equation}
In these equations $u_{1 \ast}$ and $u_{2 \ast}$ are the two limb darkening coefficients, $I(1)$ is the specific intensity at the centre of the disc and $\mu$ is defined by $\cos(\gamma)$, where $\gamma$ is the angle between the line of sight and the emergent intensity.

For both laws the two limb darkening coefficients are correlated with each other. Following \citet{Southworth08} we fix $u_{1 \ast}$ in the three bands to values which are physically appropriate for the star, and fit $u_{2 \ast}$. For the starting values we use the tables in \citet{Claret11}, interpolating to find coefficients appropriate for a star with the correct $\log g$ and $T_{\rm eff}$ (using the values for WASP-12 given in \citealt{Hebb09}). There are no tables available for the narrow band filters we use, so we assume the Sloan $g$ and $r$ tables are approximately correct for our green and red narrow band filters. We also assume a solar metallicity and a zero microturbulent velocity for the star. While these various assumptions may not be correct, they are sufficient to provide a physically appropriate starting point for our fits. 

\subsection{The model fit}
\label{sec:fitresults}

For the model fitting, we used a Markov Chain Monte Carlo (MCMC) method. This method, properly applied, potentially explores the parameter space more extensively than alternative minimisation techniques, and gives us confidence that we are finding the true minima. Our MCMC technique was described in the appendix of \citet{Copperwheat10}. As in that work we used multiple MCMC chains of $10 000+$ jumps to ensure that we obtained consistent results, although with this simplified version of the light curve model we tend to find convergence with a much smaller number of jumps. We ran two separate series of chains, using the quadratic limb darkening law for the star in one, and the non-linear Claret law in the other. $q$ is not constrained by our data and must be fixed -- we used the value determined by \citet{Hebb09}. Similarly, with only a single transit of data allowing $P$ to vary has no effect on our parameter determinations, so we fix this to the literature value. Additionally, there is a degeneracy between $R_\ast$ and $i$ which can cause a problem when these parameters are fitted simultaneously, since they are both correlated with the transit duration. We therefore use a Gaussian prior to enforce a value for $i$ close to the \citet{Hebb09} value, with a standard deviation equal to the $1 \sigma$ error. We use flat, non-informative priors for the other parameters in the model fit. The complete list of parameters that we fitted is $i$, $R_\ast$, $R_{pl}$, $t_0$, $A$, $B$ and the limb darkening coefficient $u_{2 \ast}$.

\begin{figure}
\centering
\includegraphics[angle=270,width=1.0\columnwidth]{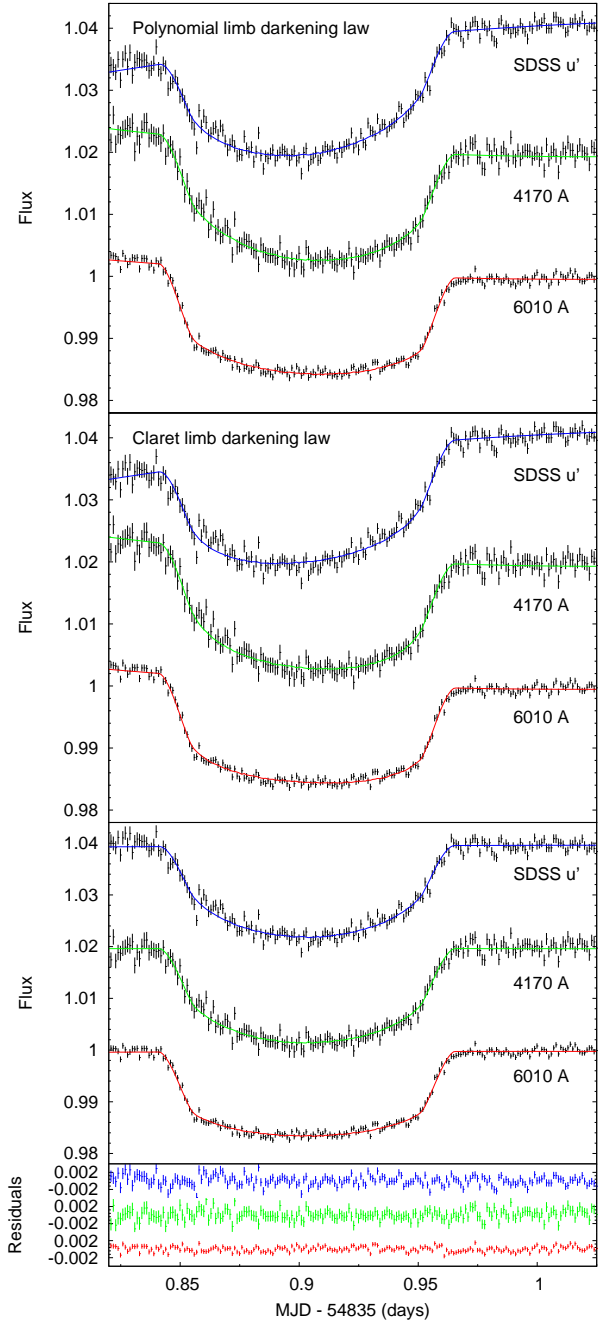}
\hfill
\caption{Light curves of WASP-12, after dividing by the averaged light curves of the comparison stars, as described in Section \ref{sec:atmos}. The data plotted here are binned such that the time between datapoints is $\sim$$85$s. The solid lines show the results of our model fits. In each panel we plot from top-to bottom the blue arm (SDSS $u$'), green arm ($4169$\AA) and red arm ($6010$\AA) data and models, with offsets of $0.02$ and $0.04$ for the green and blue arms. {\bf Top panel}: model fit when the polynomial limb darkening law is used for the star. {\bf Second panel}: model fit when the Claret law is used for the star. {\bf Third panel}: Aside from the use of comparison stars to obtain differential photometry, no other correction has been applied to flatten the light curves plotted in the first two panels. Some long-term trend remains (probably mostly airmass related), which we fit as a component in our model. Here we plot the polynomial data and model again, but with this component subtracted. {\bf Bottom panel}: Residuals from the polynomial model. The Claret model residuals are very similar.} \label{fig:lightcurve} \end{figure}

One enhancement we made here to the method described in \citet{Copperwheat10} was to fit the three bands simultaneously. This was a relatively simple modification in which the $\chi^2$ value governing the jump probability was determined from the combination of the values in the three individual bands. This modification enabled us to establish common parameters which were fit simultaneously in the three bands. We therefore determine a single, common value for $t_0$, $R_\ast$ and $i$, the parameters which are not wavelength dependent. We list our final parameter determinations in Table \ref{tab:results}. These values are the means and standard deviations from the posterior distributions. The light curves and best fit models are plotted in Figure \ref{fig:lightcurve}. 

\begin{table*}
\caption{Results from our MCMC fits. The parameters are the mass ratio $q$, the orbital inclination $i$, the stellar ($R_\ast$) and planetary ($R_{pl}$) radii scaled by the orbital separation $a$, the time of mid-transit $t_0$, the orbital period $P$, the first and second stellar limb darkening coefficients $u_{1 \ast}$ and $u_{2 \ast}$, and the two coefficients for the airmass trend $A$ and $B$. We list separately results obtained using a polynomial limb darkening law and the Claret \citep{Claret11} law. The values of $q$ and $P$ were fixed to the values given in \citet{Hebb09}. $i$, $R_\ast / a$ and $t_0$ were varied in the fit but held to the same value in all three bands.}
\label{tab:results}
\begin{center}
\begin{tabular}{lr@{\,$\pm$\,}lr@{\,$\pm$\,}lr@{\,$\pm$\,}l}
Parameter & \multicolumn{2}{l}{SDSS $u$'} & \multicolumn{2}{l}{$4169$\AA} & \multicolumn{2}{l}{$6010$\AA}\\
\hline
\multicolumn{7}{c}{Polynomial limb darkening law}\\\\
$q$		&\multicolumn{6}{c}{$0.000996$}\\
$i$ (deg)	&\multicolumn{6}{c}{$82.52 \pm 1.04$}\\
$R_\ast/a$	&\multicolumn{6}{c}{$0.3362 \pm 0.0057$}\\
$R_{pl}/a$	&$0.03933$ &$0.00105$ 			&$0.04033$ &$0.00108$ 			&$0.03970$ &$0.00094$\\
$t_0$ (days)	&\multicolumn{6}{c}{$54835.90326(13)$}\\
$P$ (days)	&\multicolumn{6}{c}{$1.091423$}\\
$u_{1 \ast}$	&\multicolumn{2}{c}{$0.5615$} 		&\multicolumn{2}{c}{$0.4647$} 		&\multicolumn{2}{c}{$0.3141$}\\
$u_{2 \ast}$	&$0.399$ &$0.061$ 			&$0.487$ &$0.075$ 			&$0.339$ &$0.063$\\
$A$		&$0.00413$ &$0.00016$ 			&$-0.00177$ &$0.00019$ 			&$-0.00128$ &$0.00009$\\
$B$		&$-0.00241$ &$0.00023$ 			&$0.00253$ &$0.00029$ 			&$0.00185$ &$0.00014$\\
\hline
\multicolumn{7}{c}{Claret limb darkening law}\\\\
$q$	&\multicolumn{6}{c}{$0.000996$}\\
$i$ (deg)	&\multicolumn{6}{c}{$83.32 \pm 1.30$}\\
$R_\ast/a$	&\multicolumn{6}{c}{$0.3302 \pm 0.0064$}\\
$R_{pl}/a$	&$0.03852$ &$0.00119$ 			&$0.03960$ &$0.00114$ 			&$0.03906$ &$0.00097$\\
$t_0$ (days)	&\multicolumn{6}{c}{$54835.90327(13)$}\\
$P$ (days)	&\multicolumn{6}{c}{$1.091423$}\\
$u_{1 \ast}$	&\multicolumn{2}{c}{$0.1632$} 		&\multicolumn{2}{c}{$0.2272$} 		&\multicolumn{2}{c}{$0.3326$}\\
$u_{2 \ast}$	&$0.719$ &$0.032$ 			&$0.626$ &$0.041$ 			&$0.300$ &$0.029$\\
$A$		&$0.00416$ &$0.00015$ 			&$-0.00173$ &$0.00020$ 			&$-0.00125$ &$0.00009$\\
$B$		&$-0.00240$ &$0.00024$ 			&$0.00251$ &$0.00030$ 			&$0.00186$ &$0.00014$\\
\hline
\end{tabular}
\end{center}
\end{table*}

\section{RESULTS}
\label{sec:parameters}

As is apparent in Figure \ref{fig:lightcurve} the model fit to the light curves is acceptable. We find a $\chi^{2}_{red}$ of $\sim$$1.2$ ($293$ $N_{\rm dof}$) in the green arm ($4169$\AA) and $\sim$$1.5$ in the other two arms. We do not find a significant difference in the quality of fit when we change limb darkening laws. The residuals show some variation, some of which might be intrinsic, but is more likely due mainly to atmospheric transmission and instrumental effects such as the flat-field `noise' which can be apparent in photometric studies at this level of precision (see, e.g., \citealt{Southworth09} for a discussion of this). Our MCMC fitting method assumes the uncertainty on each data point is Gaussian distributed and independent of all other points, so any correlated noise component (`red noise') will impact the reliability of our quoted uncertainties. Our residuals do not show any strong evidence for such a correlated component. In particular, we note that there are no apparent long-term trends in the residuals, and the model seems to fit the transit data (and the ingress/egress features) just as well as the out-of-transit data. The absence of any trends in the residual is significant: as we discussed earlier the target light curves were not flattened before fitting other than by dividing through by the average of the comparison light curves. It is clear in the data (particularly the blue arm, broadband data) that some long-term trends remain in the light curves. The fact that we do not see this trend in the residuals demonstrates that the polynomial component of our model provides an adequate fit. 

Our parameter determinations are consistent with the values reported by \citet{Hebb09}. In the case of the inclination this was enforced due to the Bayesian prior constraint in our model fitting. For the planetary radius, \citet{Hebb09} find $1.79 \pm 0.09 R_J$ using $B$- and $z$-band photometry. Our determinations tend to be a little larger than this value but are consistent with it (e.g., $1.93 \pm 0.05 R_J$ and $1.90 \pm 0.05 R_J$ for the two green arm fits, using the equatorial Jupiter radius of $71\,492\,$km). The \citet{Hebb09} determination made WASP-12b the fifth largest exoplanet currently known, but our new measurement places it in competition with WASP-17b ($1.93 \pm 0.05 R_J$, \citealt{Southworth12}) for first place. Our measurement of $t_0$ shows no evidence of a transit time variation when compared to the \citet{Hebb09} ephemeris. Recently \citet{Maciejewski13} reported a variation which they attributed to a second planet on a $3.6$ day eccentric orbit. Using their ephemeris our data is still consistent with no variation, however the epoch of our observation is close to a phase of $0$ for the sinusoidal timing variation they detect, and so our data is also consistent with the presence of this second planet.

\begin{figure*}
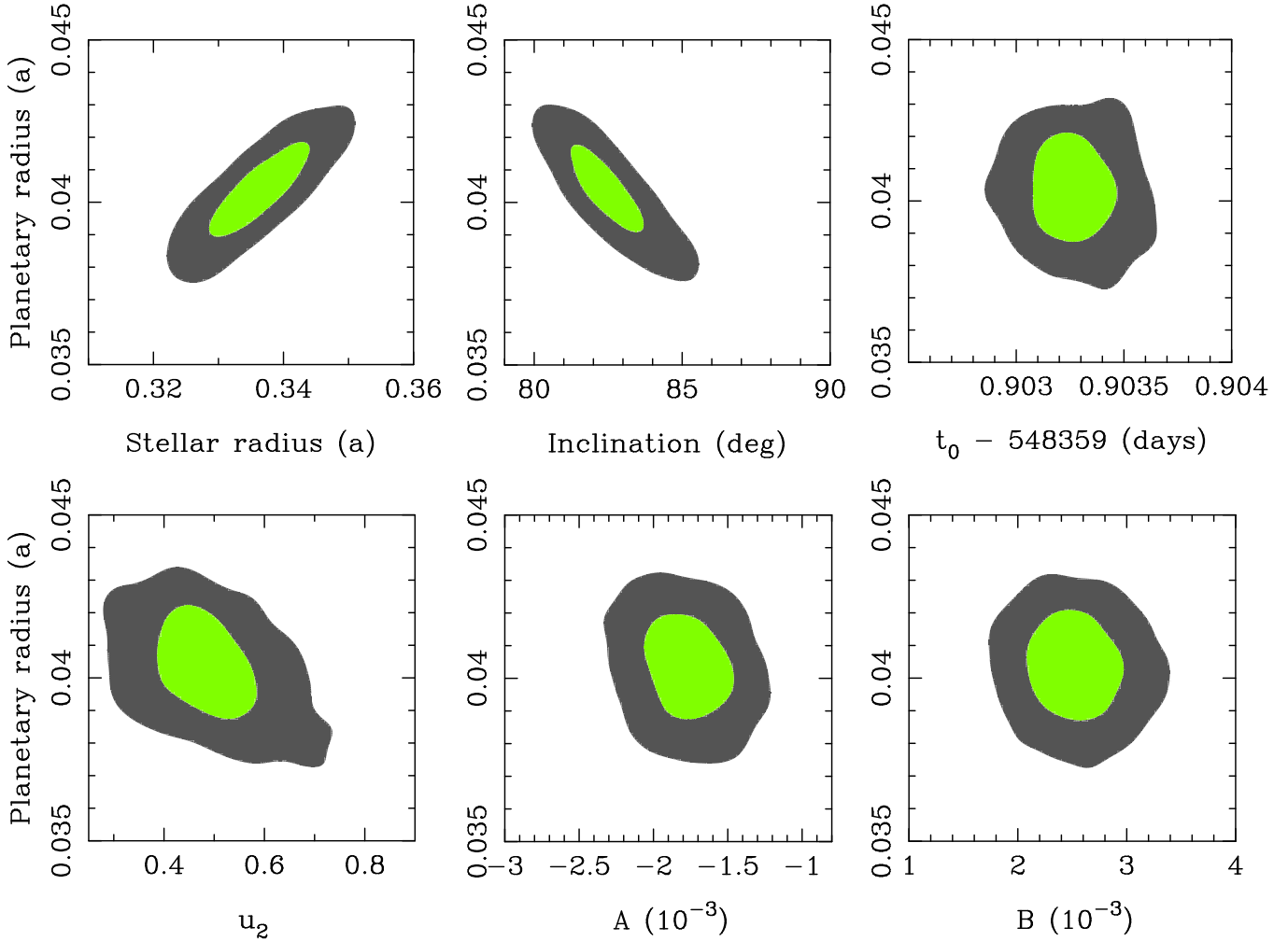

\begin{center}$
\begin{array}[t]{lll}
    \includegraphics[angle=270,width=0.35\textwidth]{rs_p.ps} & \includegraphics[angle=270,width=0.305\textwidth]{i_p.ps} & \includegraphics[angle=270,width=0.32\textwidth]{t0_p.ps}\\
    \includegraphics[angle=270,width=0.33\textwidth]{u2_p.ps} & \includegraphics[angle=270,width=0.295\textwidth]{a_p.ps} & \includegraphics[angle=270,width=0.3\textwidth]{b_p.ps}\\
\end{array}$
\end{center}
\hfill
\caption{The distribution of the results from the MCMC runs for the green arm ($4169$\AA) data when the polynomial limb darkening law is used for the star. The red and blue arm results follow similar distributions. We plot the planetary radius $R_{pl}/a$ against the other six parameters which are allowed to vary in our model fits. The green and grey regions indicate the $1\sigma$ and $3\sigma$ confidence intervals respectively.} \label{fig:mcmc_poly} \end{figure*}

\begin{figure*}
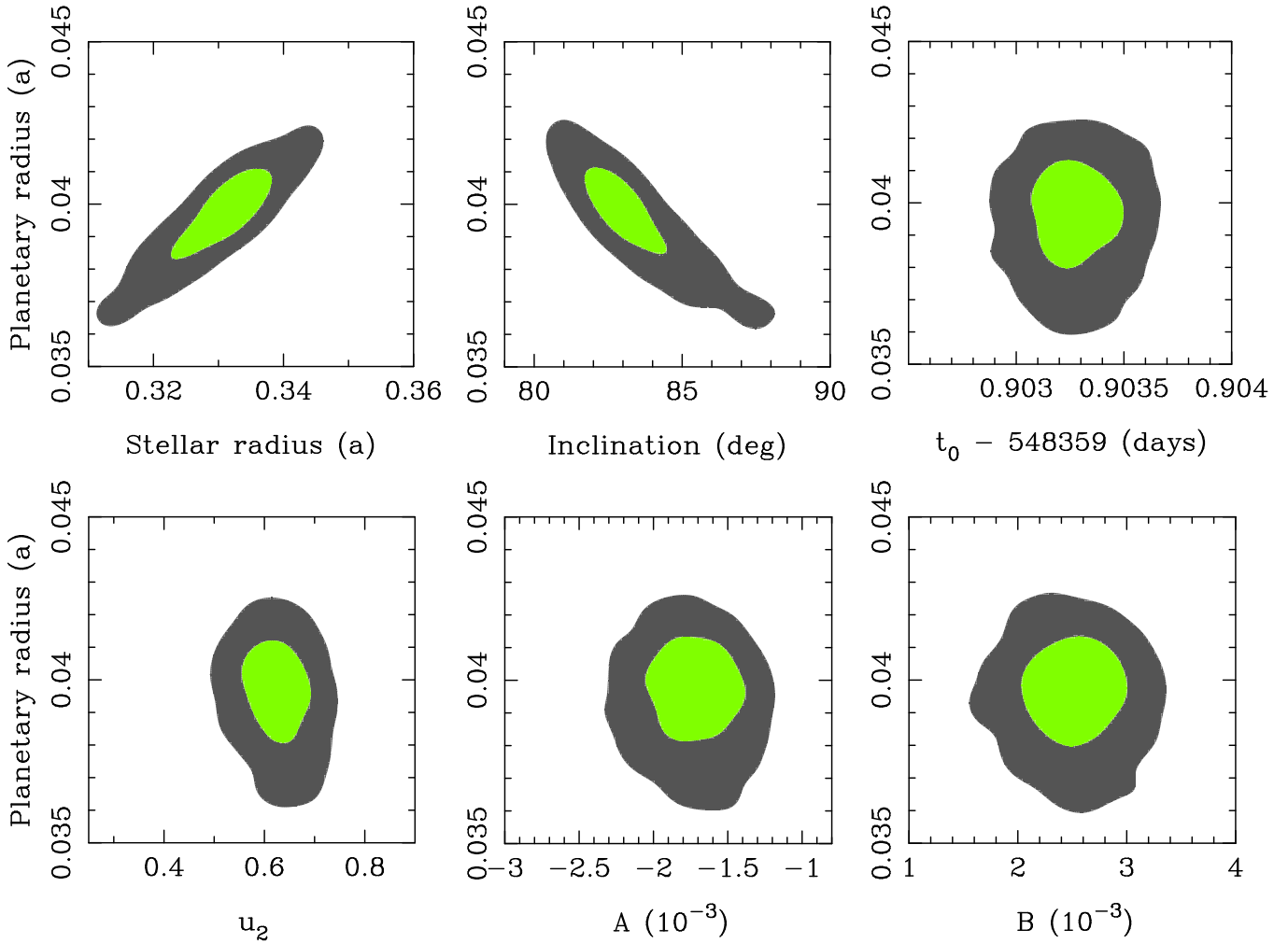

\begin{center}$
\begin{array}[t]{lll}
    \includegraphics[angle=270,width=0.35\textwidth]{rs_c.ps} & \includegraphics[angle=270,width=0.305\textwidth]{i_c.ps} & \includegraphics[angle=270,width=0.32\textwidth]{t0_c.ps}\\
    \includegraphics[angle=270,width=0.33\textwidth]{u2_c.ps} & \includegraphics[angle=270,width=0.295\textwidth]{a_c.ps} & \includegraphics[angle=270,width=0.3\textwidth]{b_c.ps}\\
\end{array}$
\end{center}
\hfill
\caption{Green arm ($4169$\AA) MCMC distributions when the Claret limb darkening law is used for the star. The red and blue arm results follow similar distributions. We plot the planetary radius $R_{pl}/a$ against the other six parameters which are allowed to vary in our model fits. The green and red regions indicate the $1\sigma$ and $3\sigma$ confidence intervals respectively.} \label{fig:mcmc_claret} \end{figure*}

In Figures \ref{fig:mcmc_poly} and \ref{fig:mcmc_claret} we plot the distributions of the results from the MCMC runs for the green arm ($4169$\AA) data, for the two different stellar limb darkening laws. We do not plot the results for the red and blue arms, but they show similar distributions. We vary seven parameters in our model fits. The key parameter is the planetary radius, so we plot this parameter against the other six in order to explore our confidence in these radius determinations.

We begin by discussing the limb darkening. We fitted our data twice, using a polynomial limb darkening law and also the Claret law. The results listed in Table \ref{tab:results} show that these two models provide consistent results, although using the polynomial law leads to a slightly higher planetary radius determination. The value of the limb darkening coefficient $u_{2 \ast}$ of course differs depending on which law is used. The fact that the two laws provide consistent results is important because we see in the distributions that $u_{2 \ast}$ can be somewhat correlated with the planetary radius, and so the fact that choice of law does not significantly bias our results increases confidence in our findings. As discussed in Section \ref{sec:limb} we chose to fix $u_{1 \ast}$ and vary $u_{2 \ast}$, since these two parameters are degenerate in our model. We explored the effect of perturbing our $u_{1 \ast}$ values from our original estimates and refitting the data, and we found that this does not have a significant effect on our results with the exception of our $u_{2 \ast}$ determination, which changes to compensate for any change in $u_{1 \ast}$. 

We also investigated how our fitted limb darkening coefficients compare to stellar model atmosphere predictions. For this we used the \citet{Kurucz79} models\footnote{http://kurucz.harvard.edu/}. We interpolated between the published grids to obtain models with a temperature and $\log g$ appropriate for Wasp-12, and convolved these grids with our filter response functions to obtain ${{I(\mu)} / {I(1)}}$ for each filter. We then used a least squares method to obtain the optimum values of $u_{2 \ast}$ for each filter and limb darkening law. As in the fits to our data, we fixed $u_{1 \ast}$ to the values listed in Table \ref{tab:results}. In general, the limb darkening coefficients predicted from these model atmosphere calculations are in reasonable agreement with the results of our model fits. When the polynomial law is used we predict values of $0.295$, $0.401$ and $0.364$ in the blue, green and red arms respectively. The blue arm prediction is $1.8\sigma$ smaller than our fitted value, but the values for the other two arms are $\sim$$1\sigma$ or less from the fitted values. For the Claret law the predictions for $u_{2 \ast}$ from the model atmospheres are $0.697$, $0.648$ and $0.374$. In this case the red arm prediction is $2.5\sigma$ greater than the determination from our fits, but the other two predictions are very close to our results.

Turning to the stellar radius and the inclination, we see in Figures \ref{fig:mcmc_poly} and \ref{fig:mcmc_claret} that these two parameters show a strong correlation with the planetary radius. They are also strongly correlated with each other, which was why we considered it necessary to use a Bayesian prior to enforce a value for the inclination close to the \citet{Hebb09} value. These parameters could have been fixed in our fits but this would result in an underestimation of the uncertainties in our planetary radius determinations. We also experimented with fixing one or the other of these parameters to the literature value, and leaving the other as a free parameter. This resulted in a determination for the free parameter which was as much as $2\sigma$ different to the literature value, with a corresponding difference in the determination of the planetary radius. However, the difference in the planetary radius determination was about the same for each light curve, with very little relative difference between the determinations in the three wavebands. This is undoubtedly due to the fact that we fit the three simultaneously. This demonstrates the power of a multi-beam camera for a study of this nature. Our conclusion is that while these parameters might introduce a systematic bias into our radius determinations, it should not affect the relative measurements and hence the conclusions we draw in Section \ref{sec:radius}.

Of the remaining three parameters, we see that there is no obvious correlation between planetary radius and the time of mid-transit ($t_0$) or the quadratic term $B$ in the airmass trend. Possibly in the polynomial law case there might be a weak correlation between the radius and the linear term $A$, which highlights the importance of including these trends as a component in the model fit so they can be properly characterised using the complete dataset.

\section{DISCUSSION}

\subsection{Planetary radius and wavelength}
\label{sec:radius}

\begin{figure*}
\centering
\includegraphics[angle=270,width=1.0\textwidth]{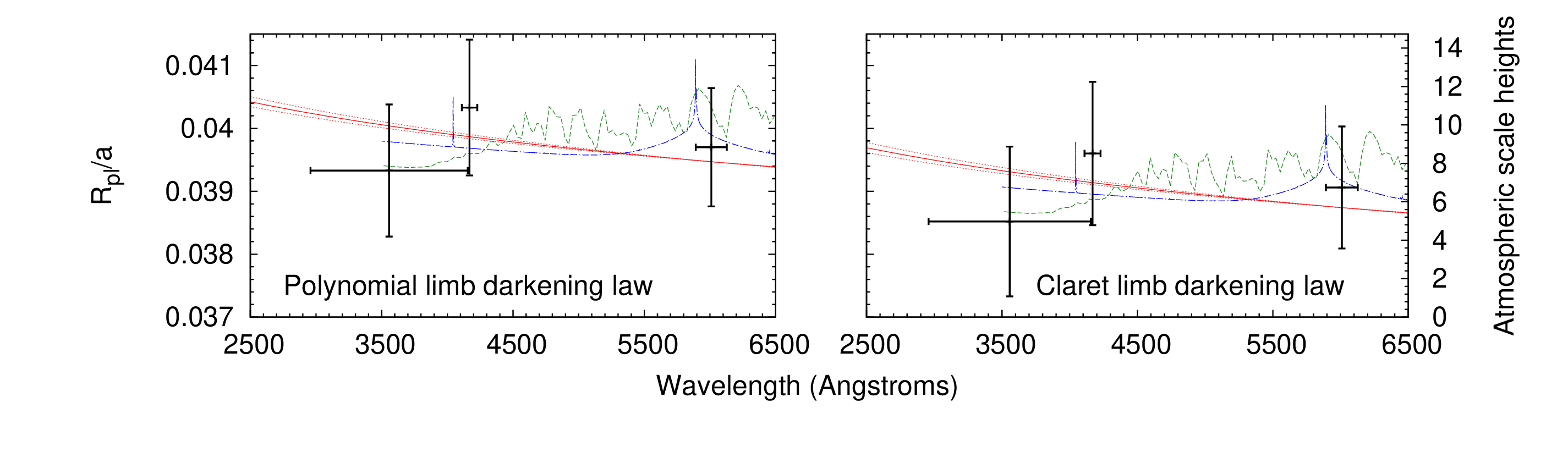}
\hfill
\caption{The planetary radius scaled by the orbital separation $a$ in each filter, as determined from our MCMC fit. We mark each measurement at the central wavelength of the filter used, with the uncertainty set to the FWHM of the filter. The right-hand axes are labelled in units of atmospheric scale heights, determined by assuming a Jupiter composition and the temperature and $\log g$ from \citet{Hebb09}. The red (solid) lines running across the plot show the predicted change in planetary radius with wavelength assuming Rayleigh scattering from a high altitude haze. This calculation also uses the scale height, and the dotted lines show the uncertainty in this determination. The other lines shows model planetary atmospheres calculated using the method described in \citet{Fortney10}. The green (dashed) line is for a 'pM-class' atmosphere in which the opacity is dominated at this wavelength range by TiO and VO, whereas the blue (dot-dashed) line is for a 'pL-class' atmosphere in which the dominant opacity sources are Na and K. An offset is applied to all three models to provide the best fit to our radius measurements.} \label{fig:wavelength} \end{figure*}

In Figure \ref{fig:wavelength} we plot our determination of the planetary radii with wavelength. As well as a radius scale in units of orbital separation $a$, we also indicate the range in terms of atmospheric scale heights, which we calculate by assuming a Jupiter composition for the planet, and the temperature and $\log g$ from \citet{Hebb09}. We plot results for the two limb darkening laws separately. The two laws provide consistent results, and we see no significant difference in planet radius with wavelength.

HST transmission spectroscopy of HD209458b \citep{Sing08,Lecavelier08} and HD189733b \citep{Pont08,Sing11} shows their spectra to be dominated by a broadband opacity source interpreted as scattering by an atmospheric haze . We calculated the predicted effect of Rayleigh scattering for WASP-12b over the wavelength range of our filters. \citet{Lecavelier08b} determined the effective transit measured altitude of a hydrostatic atmosphere as a function of wavelength. Then, by assuming a scaling law for the absorption cross section of the form $\sigma = \sigma_0 (\lambda / \lambda_0)^\alpha$, the slope of the planetary radius as a function of wavelength is given by
\begin{equation}
{{d R_{pl}}\over{d \ln \lambda}} = \alpha H,\nonumber
\end{equation}
where $H$ is the atmospheric scale height. Following \citet{Lecavelier08b}, to model Rayleigh scattering we set $\alpha = -4$. We plot the resulting slope as a red (solid) line on Figure \ref{fig:wavelength}. The dashed lines show the uncertainty in the slope, which is due to the uncertainty in the scale height determination. We apply an arbitrary offset to make the best fit of the Rayleigh scattering lines match to our radius determinations. We find that over the wavelength range we study the expected difference in planetary radius for this atmospheric model is only around $3$ -- $4$ scale heights. Hence Rayleigh scattering from a high-altitude haze is consistent with our findings. Scattering via a haze of submicrometre particles, in which $\alpha$ would be closer to zero, is equally consistent.

Other atmospheric opacity sources have been predicted, but not yet observed. Planetary atmospheres in emission have been detected with Spitzer (e.g. \citealt{Deming05}), revealing a wide range of brightness temperatures compared with expected equilibrium temperatures \citep{Harrington07}. These Spitzer results were interpreted by \citet{Fortney08} as resulting from two distinct classes of exoplanet atmosphere. The optical opacity of the hottest exoplanets (termed `pM class' planets) were predicted by their model to have atmospheres which are dominated by TiO and VO absorption bands, whereas in cooler planets (`pL-class') the TiO should have condensed out of the atmospheres and the dominant opacity sources are Na and K. We computed both models for WASP-12b, using the code described in \citet{Fortney10}. For the stellar and planetary parameters we use the \citet{Hebb09} values with the exception of the radii, for which we used our own determinations. The two models are plotted in Figure \ref{fig:wavelength}, with the green (dashed) line showing the hotter, TiO/VO dominated model, and the blue (dot-dashed) line showing the cooler, Na/K dominated model. Given the temperature of WASP-12b (estimated as $2516 \pm 36$K by \citealt{Hebb09}) we would expect the hotter model to provide a better fit to our data, however we find that both models are consistent with our radius measurements to within the uncertainties. For this planet these atmosphere models remain a viable alternative to the simple Rayleigh scattering case.

Recently \citet{Stevenson13} presented multi-object transmission spectroscopy of Wasp-12b over the wavelength range $\sim$$0.7$ -- $1$ $\mu$m obtained with the GMOS instrument mounted on the Gemini-North telescope. They find these data rule out a cloud-free $H_2$ atmosphere with no additional opacity sources. They also reanalysed {\it HST/WFC3} NIR and {\it Spitzer Space Telescope} data (originally published in \citealt{Swain12} and \citealt{Cowan12} respectively) to obtain a combined transmission spectrum over the wavelength range $0.7$ -- $5$ $\mu$m. There is evidence for a number of broad features in this spectrum, but the data is consistent with both O-rich and C-rich models, making them difficult to identify. Our data lies bluewards of all of these observations and so extends the wavelength range of the spectrum, however in order to combine datasets it is necessary to use the same orbital parameters. We therefore ran our MCMC fits again, but this time using the inclination and stellar radius from \citet{Stevenson13} as Guassian priors. \citet{Stevenson13} present their findings in terms of the transit depth $(R_{pl} /R_\ast)^2$. In our new fits we measure the transit depth to be $1.42 \pm 0.03$, $1.46 \pm 0.04$ and $1.43 \pm 0.03$ per cent in the blue, green and red arms respectively. Figure 19 of \citet{Stevenson13} shows that these values are similar to the transit depths measured in the GMOS data between $0.7$ and $1$ $\mu$m, implying that the planetary spectrum is fairly flat from this point down to $\sim$$0.3$$\mu$m.

\subsection{No evidence for an early ingress in the blue arm}
\label{sec:ingresstime}

\citet{Fossati10} presented near-ultraviolet ($2539$ -- $2811$\AA) HST transmission spectra of WASP-12b. They claimed that these data showed an early ingress of the exoplanet at these wavelengths, which they speculated was due to a disc of previously stripped material. These data were recently re-analysed in \citet{Haswell12}, and the same conclusion was reached: Table 4 of that paper reports an average phase offset of $\sim$$-0.034$ ($\sim$$0.037$ days) when the NUV measurements are compared to the optical transit. Some authors have suggested that an early ingress could be due to a highly-irradiated exoplanet overflowing its Roche lobe, leading to outflows of material from the Lagrangian points \citep{Lai10,Bisikalo12}. Alternatively, an early ingress could be caused by a bow shock of coronal material around the magnetosphere of the planet \citep{Llama11}.

Simultaneous transit observations in multiple optical wavebands provides a test of this result. The $u$'-band filter we use is redwards of the near-UV data presented by \citet{Fossati10}, but since this filter covers the Balmer jump region then it is likely that we would detect in this filter the same absorbing material that is postulated to cause the effect at near-UV wavelengths. However, we see in Figure \ref{fig:lightcurve} that there is no difference in the ingress time in our three wavebands. An offset of the order of $0.037$ days would be clearly detectable, and since we fitted the three light curves with the same time of mid-transit $t_0$, the fit in the $u$'-band would be significantly poorer if this offset were present. Additionally, the early ingress models all imply an asymmetry between the ingress and egress features. We do not observe this: our model cannot account for such an asymmetry so this would also manifest itself as a poor fit around these features. As an additional check we fitted the $u$'-band data again, separately from the other two wavebands. We find the subsequent parameter determinations are consistent with the findings from the simultaneous fit. We conclude that there is no evidence for an early ingress over this wavelength range.

\section{CONCLUSIONS}
\label{sec:conclusions}
In this paper we presented a transit light curve of the exoplanet WASP-12b. These data were obtained in January 2009 with the triple-beam camera ULTRACAM mounted on the William Herschel Telescope. We obtained simultaneous light curves in three filters: Sloan $u'$, and two longer wavelength narrow-band filters. The motivation for these observations was to attempt a photometric equivalent of the transmission spectroscopy technique which has been successfully used from both the ground and in space to characterise the atmospheres of transiting exoplanets. At the time of our observation WASP-12b was the hottest known transiting exoplanet, and so an excellent target with which to test this technique.

The main aim of our observations was to measure the planetary radius at three different wavelengths. Properly accounting for the various correlations between system parameters make a reliable determination a non-trivial task. We have been careful when analysing our data to correctly propagate all potential sources of error throughout our analysis so that the quoted uncertainties on our final radius determinations are a realistic depiction of the limitations of our data. An important part of this is that after we divide our target light curve by the light curves of the comparisons, we do not apply any additional correction to the residual trends in our data. We fit these trends as a component in our model. The reliability of our results is also greatly aided by the fact that our three light curves were obtained and fitted simultaneously. We examine the effect of perturbing some of the parameters in our model, and conclude that while the correlations between the parameters might introduce some systematic bias to our planetary radius measurements, the ratio of the radius determinations in the three wavebands is robust and reliable.  

Our data show no evidence for a difference in planetary radius over the wavelength range we study. We calculate the predicted planetary radius difference if the atmosphere were dominated by Rayleigh scattering, as in HD209458b and HD189733b. This atmosphere model is consistent with our findings. We also calculate atmosphere models for the two classes of hot Jupiter given in \citet{Fortney08}, which predicts the major opacity sources in the atmosphere are TiO and VO, or Na and K, depending on the planetary temperature. Both of these models are also consistent with our radius measurements. Our radius measurements have an average precision of $2.6$ per cent, which is close to the differences which the models predict over this wavelength range ($\sim$$1.4$ -- $2.4$ per cent). For planets with a larger atmospheric scale height, measurements with this precision would be effective at distinguishing between the models. Future applications of the transit photometry technique to WASP-12b would benefit from the measurement of multiple transits, as well as the choice of an object with more bright comparison stars in the field of view. Reducing the photon noise contribution from the comparison stars would enable us to obtain more precise determinations of the planetary radius.

We also examined our data for signs of an early ingress at blue wavelengths, which has been reported for this system in the near-UV using HST transmission spectra. All three of our light curves show an identical time of ingress and egress.

\section*{ACKNOWLEDGEMENTS}
CMC, PJW and TRM acknowledge the support of grant ST/F002599/1 from the Science and Technology Facilities Council (STFC). ULTRACAM, VSD and SPL are supported by STFC grants PP/D002370/1 and PP/E001777/1. The results presented in this paper are based on observations made with the William Herschel Telescope operated on the island of La Palma by the Isaac Newton Group in the Spanish Observatorio del Roque de los Muchachos of the Institutio de Astrofisica de Canarias. This research has made use of NASA's Astrophysics Data System Bibliographic Services and the SIMBAD data base, operated at CDS, Strasbourg, France. We thank the anonymous referee for their helpful comments which led to a number of improvements to this paper.

\bibliography{wasp12}

\end{document}